\documentclass[preprint,preprintnumbers,amsmath,amssymb,floatfix]{revtex4-1}

\usepackage{graphicx}
\usepackage{dcolumn}
\usepackage{bm}
\usepackage{epsfig}

\begin{document}

\centerline{\normalsize DESY 16--013 \hfill ISSN 0418--9833}
\centerline{\normalsize MaPhy-AvH/2016-08 \hfill}

\vspace*{1cm}

\title{Production of massless bottom jets in $p\bar{p}$ and $pp$ collisions at 
next-to-leading order of QCD}

\author{Isabella Bierenbaum}

\email{bierenbaum@math.hu-berlin.de}

\affiliation{Institut f\"ur Physik, Humboldt-Universit\"at zu
  Berlin, Unter den Linden 6, 10099 Berlin, Germany}

\author{Gustav Kramer}

\email{gustav.kramer@desy.de}

\affiliation{{II.} Institut f\"ur Theoretische Physik,
Universit\"at Hamburg, Luruper Chaussee 149, 22761 Hamburg, Germany}

\date{\today}

\begin{abstract}

We present predictions for the inclusive production of bottom jets in
proton-antiproton collisions at 1.96 TeV and proton-proton collisions
at 7 TeV. The bottom quark is considered massless.  In this scheme, we
find that at small transverse momentum ($p_T$) the ratio of the
next-to-leading order to the leading-order cross section (K factor) is
smaller than one. It increases with increasing $p_T$ and approaches
one at larger $p_T$ at a value depending essentially on the choice of
the renormalization scale. Adding non-perturbative corrections
obtained from PYTHIA Monte Carlo calculations leads to reasonable
agreement with experimental b-jet cross sections obtained by the CDF
and the CMS collaborations.

\end{abstract}

\maketitle
\thispagestyle{empty}

\section{Introduction}

The cross section for producing jets in proton-proton collisions at
the Large Hadron Collider (LHC) constitutes an important testing
ground of perturbative Quantum Chromodynamics (QCD) and offers
information for the proton parton distribution functions (PDFs). So
far, all predictions for LHC experiments, and also for Tevatron
experiments, have been found in very good agreement with the
measurements. In these predictions, the contributions of heavy quarks
(charm and bottom) are included. Since they are usually only a few
percent of the total jet production cross sections and considering the
experimental error, it is not clear, whether the heavy quark jets are
satisfactorily described by perturbative QCD (pQCD) calculations.

To test these heavy quark contributions, it is necessary to compare to
cross sections for the production of particular heavy quark jets. Such
measurements have been done recently by the CDF collaboration \cite{1}
at the Tevatron and the CMS \cite{2} and ATLAS \cite{3} collaborations
at the LHC. The CDF measurement of the inclusive b-jet production
cross section in $p\bar{p}$ collisions is performed for jets with
rapidity $y$ in the range $|y| \leq 0.7$ and for transverse momenta
$p_T$ in the region $38 < p_T < 400$ GeV, for events, in which the
decay vertex of the b-hadron is directly reconstructed. For this, the
cone-based-iterative MidPoint algorithm is applied. This algorithm is
used for jet reconstruction in the $ y-\phi$ space, assuming a cone
radius of 0.7. Further details on b-jet identification through
secondary vertex reconstruction can be found in Ref.~\cite{1}.

As a first step in a perturbative approach, one can, roughly speaking,
describe the data by a leading order (LO) calculation plus parton
shower plus hadronization, where the latter two are non-perturbative
effects. In this sense, next-to-leading order (NLO) calculations are
part of the parton shower, which has to be treated accordingly, when
higher-order calculations are explicitly included and multiplicative
factors for the correct treatment of the non-perturbative effects have
to be identified.

The measured b-jet production cross section from CDF is compared to
the leading-order PYTHIA-TUNE A Monte Carlo \cite{21} program
predictions with the CTEQ5L PDF, finding reasonably good agreement
between data and PYTHIA predictions \cite{1}, and to NLO pQCD
predictions by Frixione and Mangano \cite{4}. This theoretical NLO
prediction is based on the massive quark scheme or fixed flavor number
scheme (FFNS), in which bottom quarks appear in the final state only,
but not as partons in the initial state. Here, the comparison is done
applying a b-quark pole mass $m_b=4.75$ GeV, the CTEQ6M PDF,
renormalization and factorization scales $\mu_R$ and $\mu_F$ set to
$\mu_0 = \sqrt{p_T^2+m_b^2}/2$, and with a cone-based algorithm using
cone size $R=0.7$ and $R_{sep}= 1.3$. Furthermore, it includes an
additional correction factor for non-perturbative contributions, i.e.,
contributions from underlying event and hadronization processes equal
to 1.2 at low $p_T$ and equal to 1.0 for $p_T > 140$ GeV. The data and
theoretical cross sections are found in approximate agreement, most of
the data points lie above the theoretical predictions by nearly
50$\%$.

At the LHC, two collaborations, CMS \cite{2} and ATLAS \cite{3}, have
measured inclusive b-jet cross sections at 7 TeV center-of-mass (c.m.)
energy. In this work, we shall concentrate on the comparison with the
CMS data \cite{2}. The jet transverse momentum in the cross sections
presented by CMS lies between $18 < p_T < 200$ GeV for several
rapidity intervals in the range $0.0< |y| < 2.2$. To obtain the
bottom jets, the anti-$k_T$ algorithm \cite{5} with R=0.5 is used
experimentally, as well as in the theoretical NLO calculations
\cite{4}. In the CMS experiment, b-jets are identified by finding the
secondary decay vertex of the b-hadrons. The measured cross sections
have been compared to the PYTHIA Monte Carlo generator \cite{6}, and
to the MC@NLO \cite{7,8} Monte Carlo generator which is essentially
based on the FFNS NLO cross sections \cite{4} with the subsequent
Monte Carlo HERWIG generator \cite{9} which contains shower and
hadronic corrections. The MC@NLO predictions are below the data in the
central region ($|y| < 1.0$) for low $p_T$ by approximately a factor
of 1.5 to 2.0. The predictions from the PYTHIA generator agree with
the data at high $p_T$, but overestimate the cross section in the
$p_T$ region below 50 GeV.

It is clear that the PYTHIA and also the MC@NLO prediction depend on
the amount of hadron corrections contained in these two Monte Carlo
generators and it is not known which amount originates from the
underlying leading-order or next-to-leading order hard perturbative
QCD cross sections, the parton shower and the hadronic corrections.

It is the purpose of this work to find out the NLO perturbative QCD
cross section for inclusive b-jet production at the Tevatron and the
LHC center-of-mass energies and to supplement the NLO predictions with
estimates of the hadronic corrections. In our calculation, the bottom
quark is considered massless, like all other quarks u, d, s and c. So
the calculations are performed in the zero-mass variable-flavor-number
scheme (ZM-VFNS) as in our work on charm jets \cite{10}. Such a
framework has also been considered previously by Banfi et
al. \cite{14,15}. They started from the NLOJET \cite{16,17} program,
which is an alternative to calculate jet cross sections in
hadron-hadron collisions, where one sums over all flavors of outgoing
partons. This program was modified in such a way that only b quarks
(or c quarks) appear in the final state. In addition, they changed the
jet algorithm for b quarks. Unfortunately, this new jet algorithm for
heavy quarks has not been used in the analysis of the experimental
data yet. It is not clear whether this can be realized.

In Sec. 2, we shall describe the theoretical framework and outline the
PDF input for the initial state. Section 3 contains our results for
the bottom-jet cross section. In this section, we also show a
comparison with the single-inclusive jet cross section measured by the
CDF collaboration at the Tevatron.  Conclusions and summary are
presented in Sec. 4.

\section{Theoretical Framework and PDF Input}

As in our publication on charm jets \cite{10}, we rely on previous
work on dijet production in the reaction $\gamma + p \rightarrow jet +
X$ \cite{11,12}, in which cross sections for inclusive one-jet and
two-jet production up to NLO for both, the direct and the resolved
contributions are calculated. The resolved part of this cross section
routine can be used for $p\bar{p}$ or $pp$ collisions replacing the
photon PDF by the (anti)proton PDF. The routine \cite{11,12} contains
quarks of all flavours up to and including the bottom quark, as well
as the gluon. This routine has been modified in such a way that at
least one bottom quark appears in the final state, in the same way as
we did for the charm quark for charm jets in \cite{10}.

The routine \cite{11,12} is written for massless quarks, i.e. also the bottom
quark is considered massless. This is justified as long as the transverse
momentum $p_T$ of the produced jets is large enough, i.e. for $p_T^2 \gg
m_b^2$.

For our prediction of the inclusive b-jet cross section, we employ the
MSTW2008NLO \cite{13} PDF (central value) of the Durham
collaboration. The chosen asymptotic scale parameter
$\Lambda_{\overline{MS}}^{(5)}=0.262$ GeV corresponds to
$\alpha_s^{(5)}(m_Z)=0.118$. We choose the renormalization scale
$\mu_R=\xi_R p_T$ and the factorization scale $\mu_F= \xi_F p_T$,
where $p_T$ is the largest transverse momentum of the two or three
final state jets. $\xi_R$ and $\xi_F$ are dimensionless scale factors,
which are varied around $\xi_R=\xi_F=1$ in a manner to be specified
later. The center-of-mass energy of the proton-proton collisions is
taken as $\sqrt{s}=7$ TeV as for the data of CMS \cite{2}, and as
$\sqrt{s}=1.96$ TeV for the $p\bar{p}$ collisions as for the data of
CDF \cite{1}.

The experimental data of CDF have been read off from the publications
in Ref.~\cite{1}. Since the bin size in $p_T$ was not specified there,
we did our calculation with a bin size of $\Delta p_T = 10$ GeV and
plotted the CDF data for fixed $p_T$ values. The bin size in $p_T$ for
the CMS data is taken from Ref.~\cite{18}. The CMS data \cite{2}
appear for five bin sizes in $|y|$ between $0 < |y| < 2.2$. In our
comparison with these data, we shall limit ourselves to the two most
central bin sizes: $0 < |y| < 0.5$ and $0.5 < |y| < 1.0$, where the
discrepancy with the MC@NLO \cite{7,8} predictions is the largest.

\section{Results}

\subsection{Comparison with CDF data}

We start with the predictions for and the comparison to the inclusive
b-jet production data obtained by the CDF Collaboration at the
Tevatron \cite{1}. In order to check our program, we have first
calculated the cross section $d\sigma/dp_Tdy$ for $p+\bar{p}
\rightarrow single~jet + X$ in the $p_T$ range $54 < p_T < 700$ GeV
and with rapidity $|y| < 0.1$ in $p_T$ bins as chosen by CDF \cite{19}
for their measurement.  Correction factors that approximately account
for non-perturbative contributions from the underlying event and
fragmentation of partons into hadrons as given in Ref.~\cite{19} were
multiplied to the theoretical results. These non-perturbative
correction factors are estimated using the PYTHIA-TUNE A Monte Carlo
routine \cite{20,21}. The correction factors decrease with increasing
$p_T$ and lie below 1.2. The jets in the CDF analysis, as in our
calculations, are defined with the $k_T$ algorithm \cite{22,23} with
the radius R=0.7.

Our results are shown in Fig.~1, and compared to the CDF data
\cite{19}. The agreement is satisfactory. Deviations occur for small
$p_T$ and for the largest $p_T$ bin. The scales are
$\mu_R=\mu_F=p_T/2$ and the theoretical error in our calculations has
been estimated by varying $\mu_R=\mu_F$ by a factor of 2 up and down
as usual. As seen in Fig.~1, most of the data points lie inside the
theoretical error band.  Our results can also be compared to the
theoretical curves in Ref.~\cite{19} which have been calculated with
the JETRAD routine \cite{24} and the PDF set CTEQ6.1M \cite{25}
instead of the more modern PDF CT10 \cite{26} used in our work.  These
different choices of PDFs may explain small differences in the NLO
predictions in \cite{19} as compared to our calculation.
\begin{figure*}
\includegraphics[width=7.5cm]{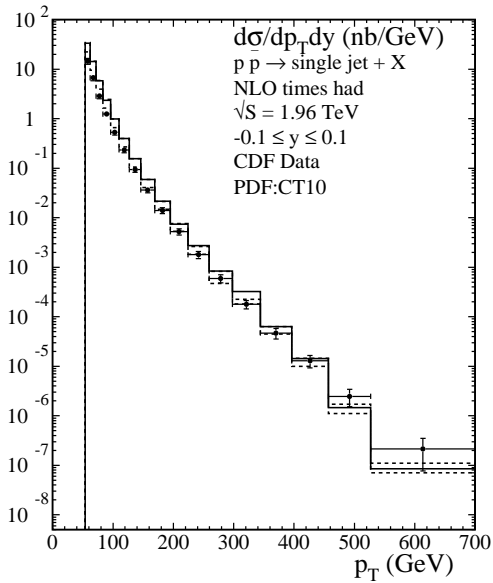}
\caption{\label{fig:1} Single-inclusive jet cross sections
  $d\sigma/dp_Tdy$ as a function of $p_T$ compared to the data from
  CDF \cite{19}. The NLO theoretical predictions are corrected for
  non-perturbative effects via multiplicative factors. The theoretical
  error is obtained by independent scale variations given as the
  dashed curves. The solid curve indicates the default scale choice.}
\end{figure*}

The results for the inclusive b-jet cross sections at $\sqrt{s}=1.96$
TeV are shown in Fig.~2 in LO pQCD and as K factors, where K is the
ratio of the NLO to the LO cross section $K =
(d\sigma/dp_Tdy)_{NLO}/(d\sigma/dp_Tdy)_{LO}$ as a function of
$p_T$. The scales for the LO cross sections are chosen as in \cite{1},
namely $\xi_R=\xi_F=0.5$ (default, full line), $\xi_R=\xi_F=0.25$
(upper broken line) and $\xi_R=\xi_F= 1.0$ (lower broken line). In
Fig.~2, we have also plotted the data for the inclusive b-jet cross
section as measured by CDF \cite{1}. These data in numerical form have
been read from the publication in \cite{1}. As we can see, the
corresponding LO cross sections are smaller than the experimental data
by approximately a factor of 2 for low $p_T$ and 4 for the largest
$p_T$ bin. The LO cross sections are multiplied with the hadronization
correction factor 1.2 as reported in \cite{1}.

If as usual, i.e. for jet cross sections of all flavors, the NLO
corrections would increase the b-jet cross sections by factors 2 to 3,
we would have approximate agreement with the data. Unfortunately, this
is not the case as we can see from Fig.~2 in the right frame. The K
factors are smaller than 1 for small $p_T$ and approach values of 1
for larger values of $p_T$ depending on the choice of scales. For the
scale choice $\xi_R=2.0$, $\xi_F=1.0$ (full line) this occurs at
$p_T=250$ GeV and for the larger scale $\xi_R=4.0$, $\xi_F=2.0$ at
$p_T=150$ GeV (upper broken line). For the usual scale choice
$\xi_R=\xi_F=1.0$ the K factor is smaller than 1 in the whole
considered $p_t$ range up to $p_T=300$ GeV (lower broken line).
\begin{figure*}
\includegraphics[width=7.5cm]{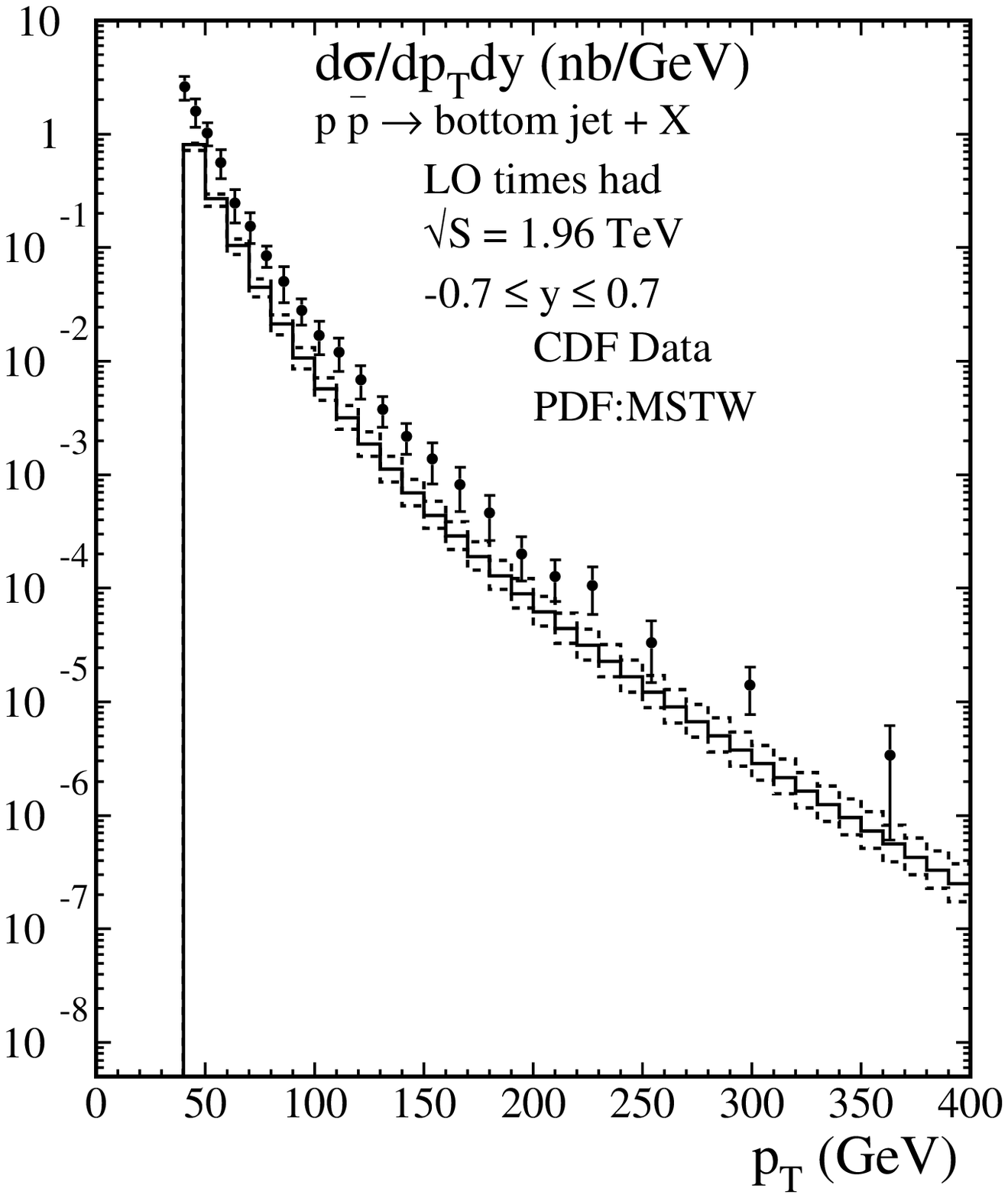}
\includegraphics[width=7.5cm]{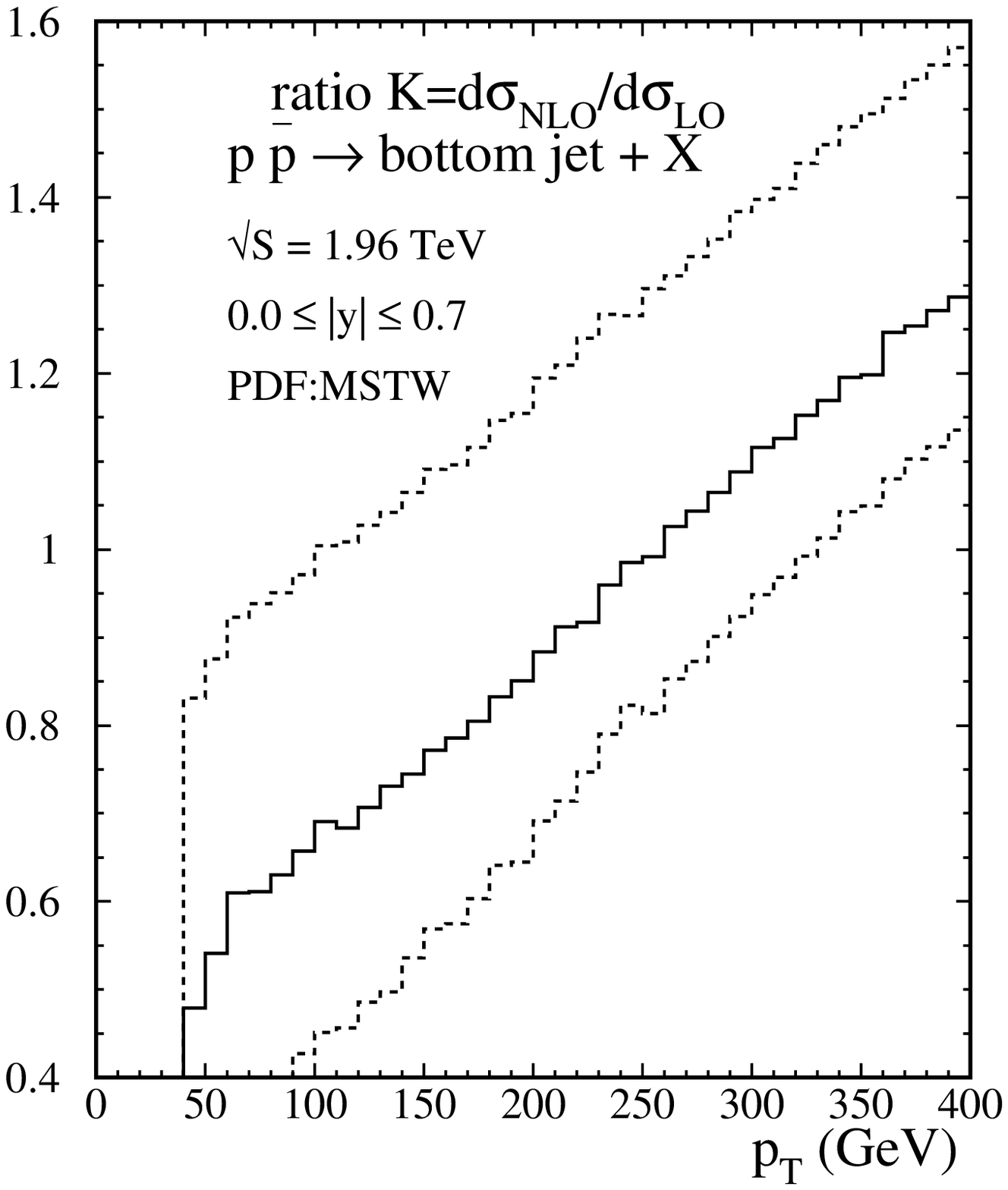}
\caption{\label{fig:2} Left-hand side: single-inclusive b-jet cross
  section in LO as a function of $p_T$ for the rapidity region $0.0
  \leq |y| \leq 0.7$ compared to CDF data \cite{1}. The LO theoretical
  predictions are corrected by non-perturbative effects via a
  multiplicative factor. The theoretical error (dashed lines) is
  obtained by scale variation as given in the text. The solid line
  indicates the default scale choice. Right-hand side: Ratio of
  single-inclusive b-jet cross section in NLO and LO for the rapidity
  region $0.0 \leq |y| \leq 0.7$ as a function of $p_T$, for three
  scale choices as given in the text.}
\end{figure*}
The reason for the K factor to be smaller than 1 at small $p_T$ lies
in the fact that at these $p_T$ values the jet cross section for all
flavors as in Fig.~1 has large contributions from the hard scattering
cross sections for light quarks as for example $gg \to q \bar{q} g$,
where q stands for u, d, s, c quarks.  Of course, these contributions
are missing in the b-jet cross section. At larger $p_T$, these light
quark contributions are presumably less important.

These results can be compared to the b-jet cross sections obtained in
the FFNS, which can be found in Ref.~\cite{1} and which have been
calculated with the NLO routine in Ref.~\cite{4}. These theoretical
results show the expected scale dependence, i.e. smaller scales for
$\mu_R=\mu_F$ lead to larger cross sections. The scale $\mu_R=\mu_0/2$
is chosen as the default scale, $\mu_R=\mu_0/4$ for the maximal and
$\mu_R=\mu_0$ for the minimal cross section. In our case with
massless b quarks, it is just the opposite, increasing the scale
$\mu_R$ leads to larger b-jet cross sections, since the NLO
corrections are reduced with increasing $\mu_R$. Second, the agreement
between theory and data in Ref.~\cite{1} is best for $\mu_R=\mu_0/4$
in the range of $p_T > 80$ GeV and for $\mu_R=\mu_0/2$ if $p_T < 80$
GeV, whereas in our case for all considered scales the data lie above
the predicted cross sections.

Our NLO cross sections are shown in Fig.~3, now for the scale choices:
$\xi_R=4.0$, $\xi_F=2.0$ (full line), $\xi_R=8.0$, $\xi_F=4.0$ (upper
broken line) and $\xi_R=2.0$, $\xi_F=1.0$ (lower broken line). The
first and the last of these scale choices agree with those made for
the K factor calculation in Fig.~2 (right frame). The choice of a
larger renormalization scale, of course, has the effect that the NLO
corrections to the cross section are reduced. At large $p_T$, the
b-jet cross section for all three scales is almost the same as in the
LO approximation (see Fig.~2, left frame) as to be expected from the K
factors in Fig.~2 (right frame). For these results, the jet algorithm
is the cone algorithm with $R=0.7$ and $R_{sep}=1.3$ which is
considered best to represent the cone choice in the experimental
analysis \cite{4}. As PDF we have chosen the MSTW2008NLO version
\cite{13}. The NLO curves in Fig.~3 are compared to the data for
inclusive b-jet cross sections as measured by CDF \cite{1}. The
agreement is similar as for the LO cross sections (see Fig.~2
left). 

The cross-section calculations in the FFNS and in the ZM-VFNS behave
quite differently, which follows already from the different scale
dependence. The reason for this though is not known yet and would
require a more detailed investigation, which is beyond this work. One
possible reason might be that in our approach with massless bottom
quarks, there are additional NLO contributions to the b-jet cross
section originating from the b-quark PDF of the proton (antiproton)
not present in the FFNS, since there it would amount to a NNLO
contribution.

The unusual behaviour of the theoretical NLO prediction in comparison 
with the CDF data might be explained by missing non-perturbative hadronic and
underlying event contributions. 
\begin{figure*}
\includegraphics[width=7.5cm]{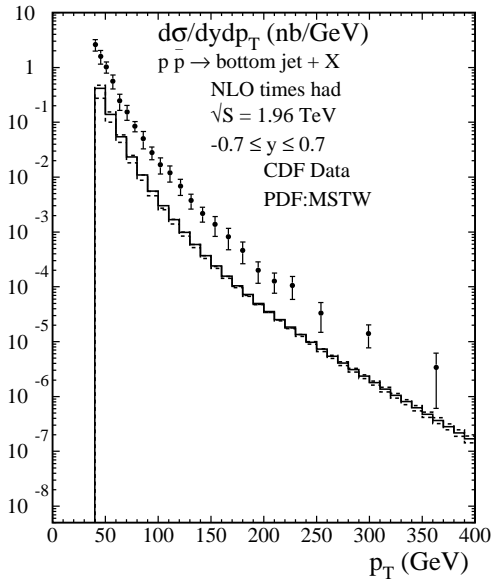}
\caption{\label{fig:3} Single-inclusive b-jet cross section in NLO as
  a function of $p_T$ for the rapidity region $0.0 \leq |y| \leq 0.7$
  compared to CDF data \cite{1}. The NLO theoretical predictions are
  corrected by non-perturbative effects via a multiplicative
  factor. The theoretical error (dashed lines) is obtained by scale
  variation as given in the text. The solid line indicates the default
  scale choice.}
\end{figure*}
The CDF data have been compared in Ref.~\cite{1} also with the
PYTHIA-TUNE A \cite{21,28} predictions and good agreement has been
found up to $p_T =200$ GeV. The PYTHIA Monte Carlo routines are based
on LO hard production matrix elements and LO PDFs. Our LO predictions
have been done with the same PDF \cite{13} as for our NLO predictions
and the results are shown in Fig.~2 left frame. The different choice
of PDFs should have only little influence in this case. We saw in that
figure that the LO cross section is nearer to the CDF data and similar
to the NLO cross section. According to Ref.~\cite{1}, the PYTHIA
prediction and the CDF data agree more or less, so that the difference
between PYTHIA and LO predictions would give a correction factor which
brings the LO prediction into agreement and the NLO prediction, which
is somewhat smaller at low $p_T$ than the LO prediction, much closer
to the data than the NLO curve plotted in Fig.~3. In other words, for
NLO, the data of CDF need a much larger non-perturbative correction
factor than the factor 1.2 which we applied in our plot of the NLO
cross section in Fig.~3. The correction factor 1.2 originates from the
same PYTHIA calculation, but with the difference that the
contributions from multiple parton interactions and string
fragmentation are subtracted from the PYTHIA result, because they are
usually considered already to be part of the NLO corrections. This
means that a correction factor that contains besides the
non-perturbative hadronization corrections and contributions from the
underlying event also the parton showers must be applied to come
closer to the experimental cross section. As is well known, this
implies some double counting since the parton showers contain some NLO
corrections which are already contained in the perturbative NLO
contributions.

We shall see in the next subsection that the same pattern also follows
from the comparison with the b-jet production cross section as
measured by CMS \cite{2}. Since we do not have the PYTHIA predictions
at our disposal, we cannot be more quantitative concerning the
comparison of the CDF data with our NLO calculations. It seems,
however, that a much larger part of the b-jet cross section originates
from the non-perturbative contribution and not just from the
perturbative NLO contribution as it is the case for the single-jet
cross section of all flavors.

At this point, we want to mention that the non-perturbative
corrections are applied to the NLO b-jet cross section in form of a
factor 1.2, taken from \cite{1} and obtained from the PYTHIA Monte
Carlo routine which is based on LO b-jet cross section with massive b
and c quarks and only u-, d-, s- and g-PDFs of the proton and
antiproton. From earlier studies we know that the b-jet and c-jet
production cross section in LO differs only at rather small $p_T
\simeq m_b, m_c$ from the massless LO cross sections. How finite
values of $m_b$ and $m_c$ influence the parton shower contributions is
not known. In addition, it is also not known how much the
non-perturbative corrections are changed, if one adds b- and c-quark
PDFs of the in-going hadrons in the LO cross section.

\subsection{Comparison with CMS data}

The CMS Collaboration has also measured the $p + p \rightarrow
single~jet +X$ cross section for all flavors in the range $18 < p_T <
1684$ GeV \cite{29}. This has been compared to our calculations using
the same routine as in this work in Fig.~1 of Ref.~\cite{10}, with the
result that good agreement between predictions and data has been
found. In the following, we show the comparison of our prediction to
the cross section measurements for single bottom production $p + p
\rightarrow bottom~jet +X$ published in Ref.~\cite{2}. We do this for
two rapidity regions $|y| \leq 0.5$ and $0.5 \leq |y| \leq 1.0$, and
for the jet transverse momenta in the range between $18 < p_T < 200$
GeV.

In Ref.~\cite{2}, the CMS data have been compared to PYTHIA predictions
\cite{6} with the result that for $p_T > 50$ GeV good agreement had been
found, whereas for $p_T < 50$ GeV the PYTHIA prediction was much larger than
the data. This also occurred for the other three rapidity ranges not
considered in this work. In addition, the CMS data have been compared to the
MC@NLO predictions. For the two lowest $|y|$-regions, the MC@NLO values are
below the data by a factor of 1.5 to 2.0 almost over the whole range of $p_T$
values.
\begin{figure*}
\includegraphics[width=7.5cm]{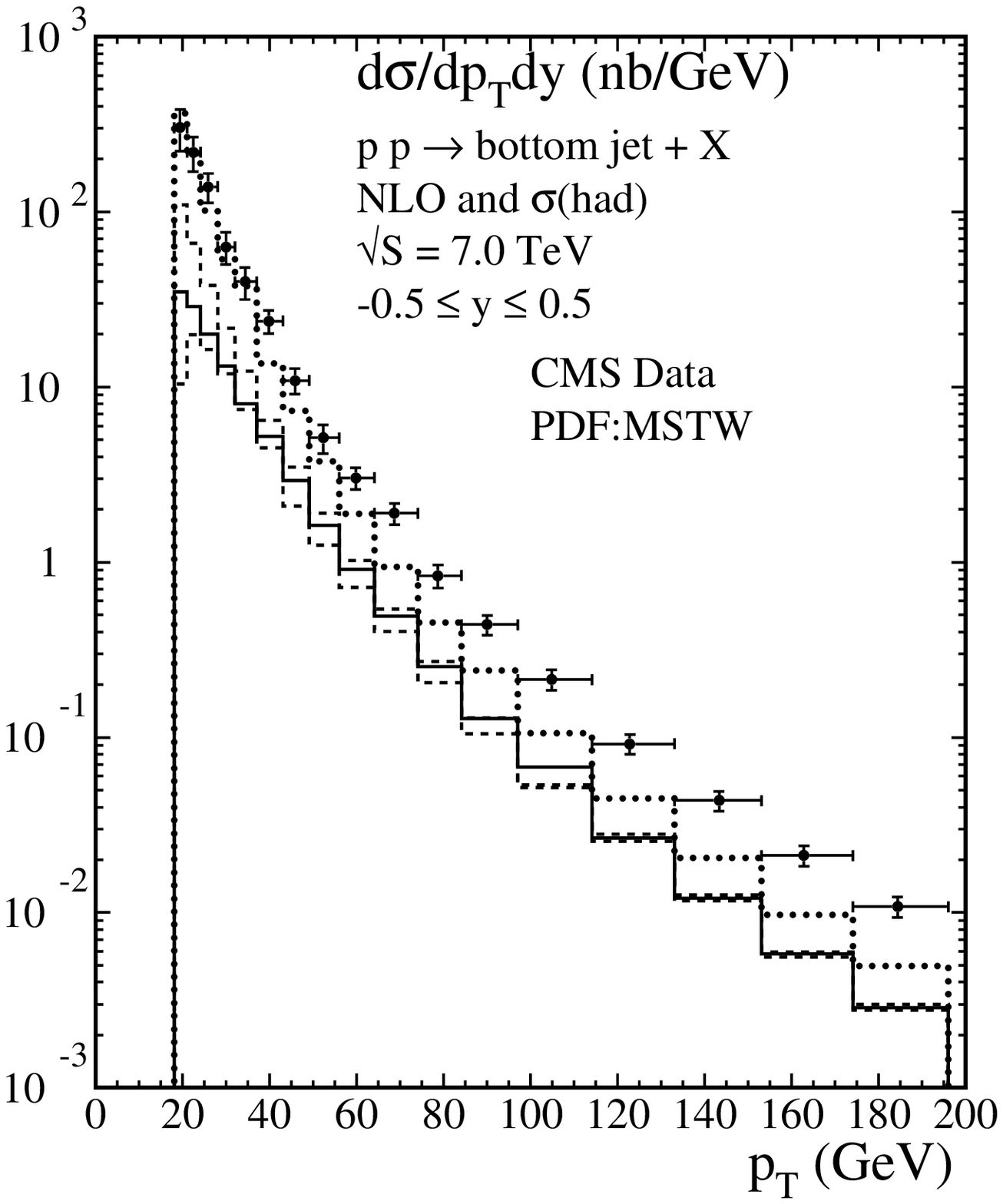}
\includegraphics[width=7.5cm]{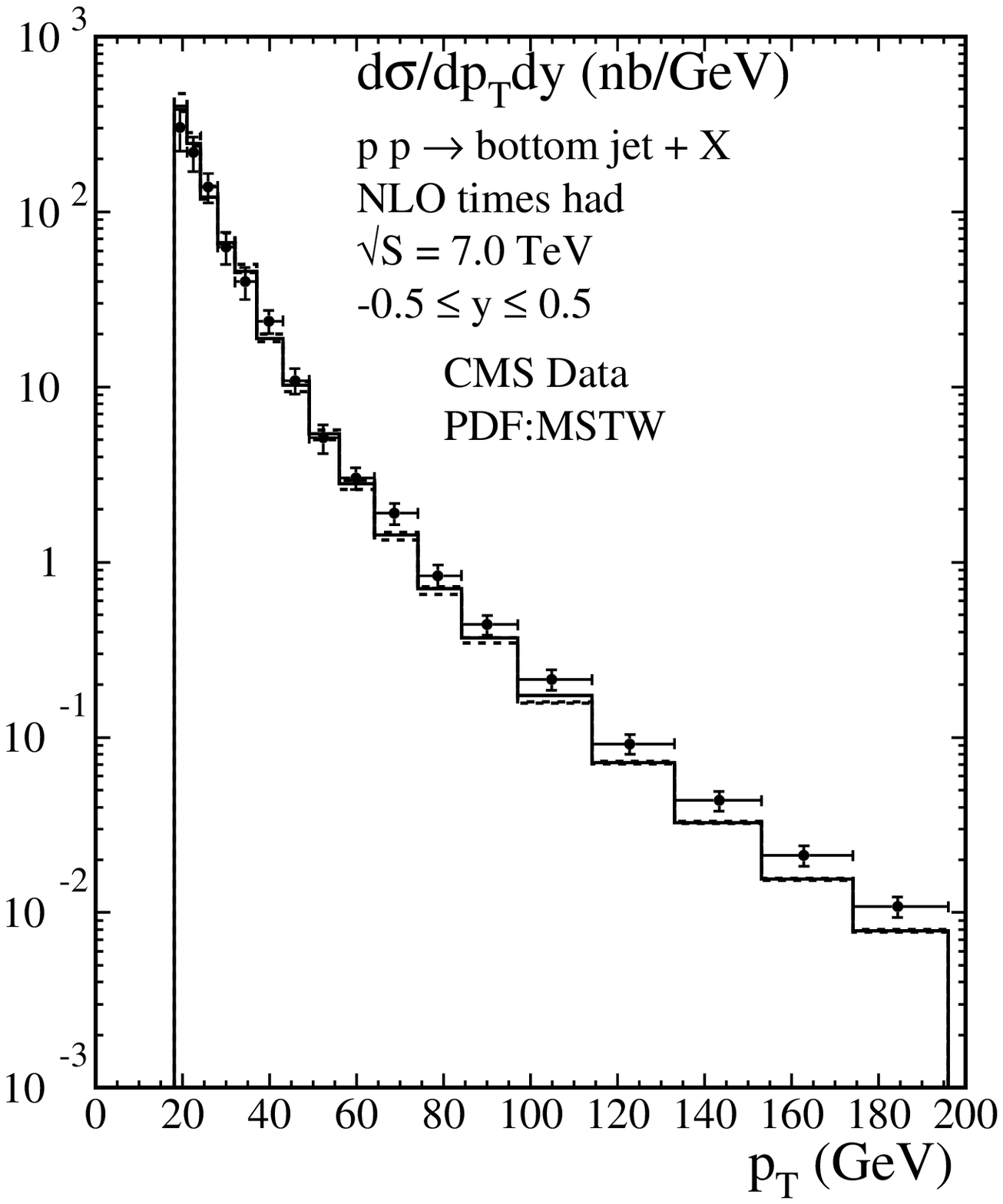}
\caption{\label{fig:4} Left-hand side: Single-inclusive b-jet cross
  section in NLO as a function of $p_T$ for the rapidity region $0.0
  \leq |y| \leq 0.5$ and compared to CMS data \cite{2}. The NLO
  theoretical prediction is not corrected by non-perturbative effects
  via multiplicative factors. The theoretical error (dashed
  histograms) is obtained by scale variation as given in the text. The
  solid histogram indicates the default scale choice. The dotted
  histogram is the PYTHIA minus LO prediction. Right-hand side:
  Single-inclusive b-jet cross section in NLO plus the PYTHIA minus LO
  prediction as a function of $p_T$ for the rapidity region $0.0 \leq
  |y| \leq 0.5$, compared to CMS data \cite{2}.}
\end{figure*}

Next, we want to see how well our NLO predictions describe the CMS
b-jet cross section data. For this purpose, we need to know the
non-perturbative corrections. As an estimate for these corrections, we
take the PYTHIA predictions made available to us in numerical form as
they were plotted in Ref.~\cite{2} by R. Kogler and M.A. Voutilainen
from the CMS Collaboration \cite{30}. From the PYTHIA cross section,
we subtracted the LO predictions as we proposed already for the
comparison with the CDF data. The PYTHIA minus LO predictions are
plotted in Fig.~4 and Fig.~5 for both rapidity regions $|y| \leq 0.5$
(dotted line in the left frame of Fig.~4) and for $0.5 \leq |y| \leq
1.0$ (dotted line in the left frame of Fig.~5).  As can be seen form
these two figures, for $p_T \leq 35$ GeV the CMS experimental data
almost agree with these non-perturbative contributions. For $p_T > 35$
GeV the data points deviate from the dotted histogram and at the
largest $p_T$ bin, the dotted histogram yields approximately $50\%$ of
the experimental cross section for both $|y|$ regions.

In Fig.~4 (left frame) and Fig.~5 (left frame), we have also plotted
our prediction for the NLO cross section (full line histogram)
together with the scale variation (dashed line histograms). These NLO
cross sections are obtained with the scale factors $\xi_R=4.0$,
$\xi_F=1.0$ (default prediction), $\xi_R=8.0$, $\xi_F=2.0$ (upper
broken line) and $\xi_R=2.0$, $\xi_F=1.0$ (lower broken line). The
value $\xi_R=4.0$ for the default prediction has been chosen in order
to reduce the coupling constant to obtain a reasonably large NLO cross
section for the smaller $p_T$, where the NLO cross section is still
smaller than the LO prediction, giving rise to a K factor smaller than
one, as we observed already for the prediction of the CDF data. At
very small $p_T$, our prediction in NLO is approximately a factor 10
smaller than the data points so that the non-perturbative part alone
produces the b-jet cross section.  For the larger $p_T$ this factor
reduces to 3. This means that for all $p_T$ our NLO prediction is
below the measured cross section data. The scale dependencies of the
NLO prediction decrease towards larger $p_T$ and have, except for the
first five $p_T$ bins, no large effect.

In Fig.~4 (right frame) and Fig.~5 (right frame), the NLO cross
sections and the “non-perturbative” contributions PYTHIA minus LO
cross section have been added and compared to the CMS data. We find
reasonable agreement between this prediction and the data inside the
experimental errors up to approximately $p_T \simeq 100$ GeV. For
larger $p_T$, the difference between prediction and data increases. But
even at the largest $p_T$ bin, the difference between the predicted
and the measured cross section is not more than $20\%$, while the
measured cross section is throughout larger than the theoretical cross
section.
\begin{figure*}
\includegraphics[width=7.5cm]{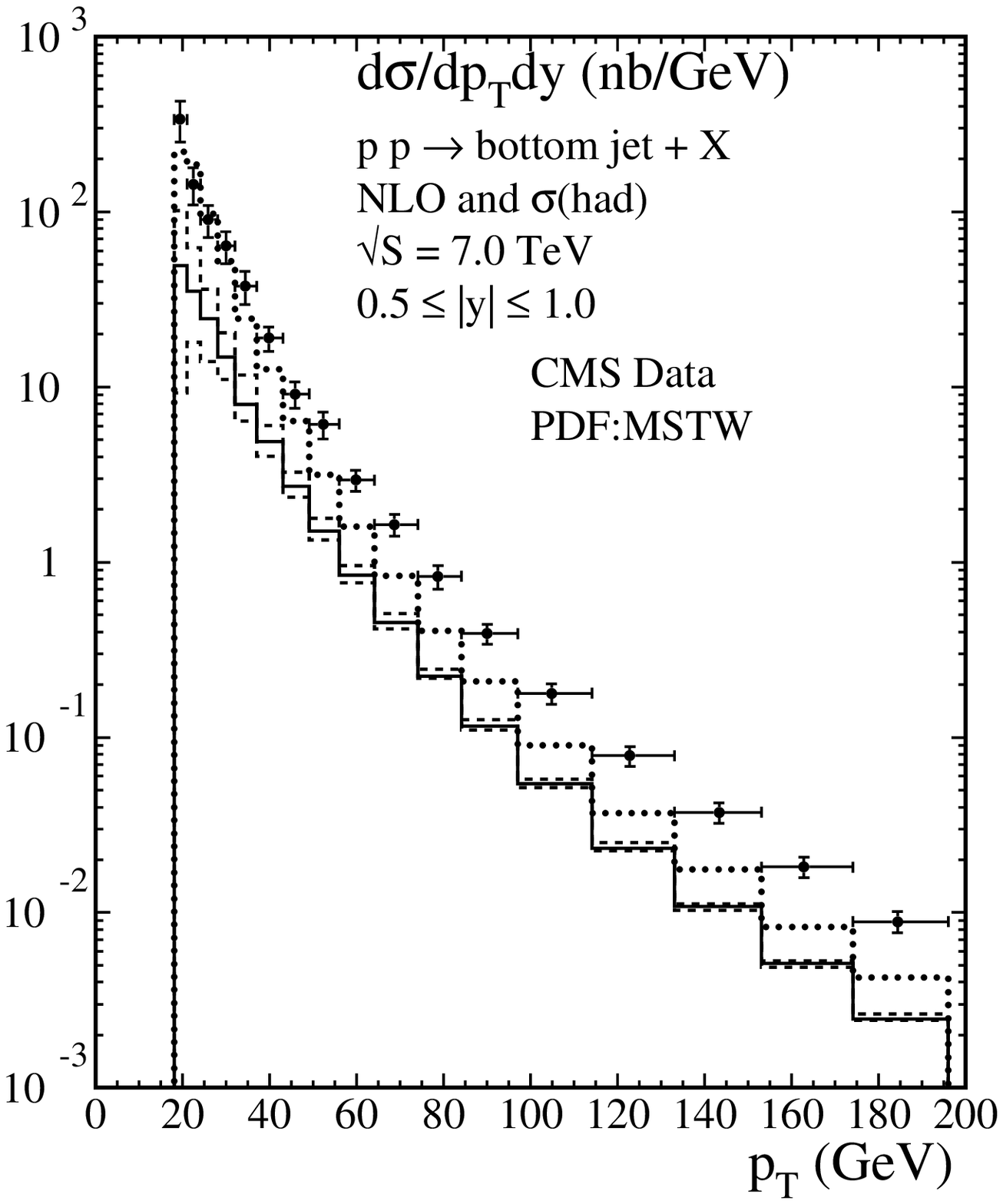}
\includegraphics[width=7.5cm]{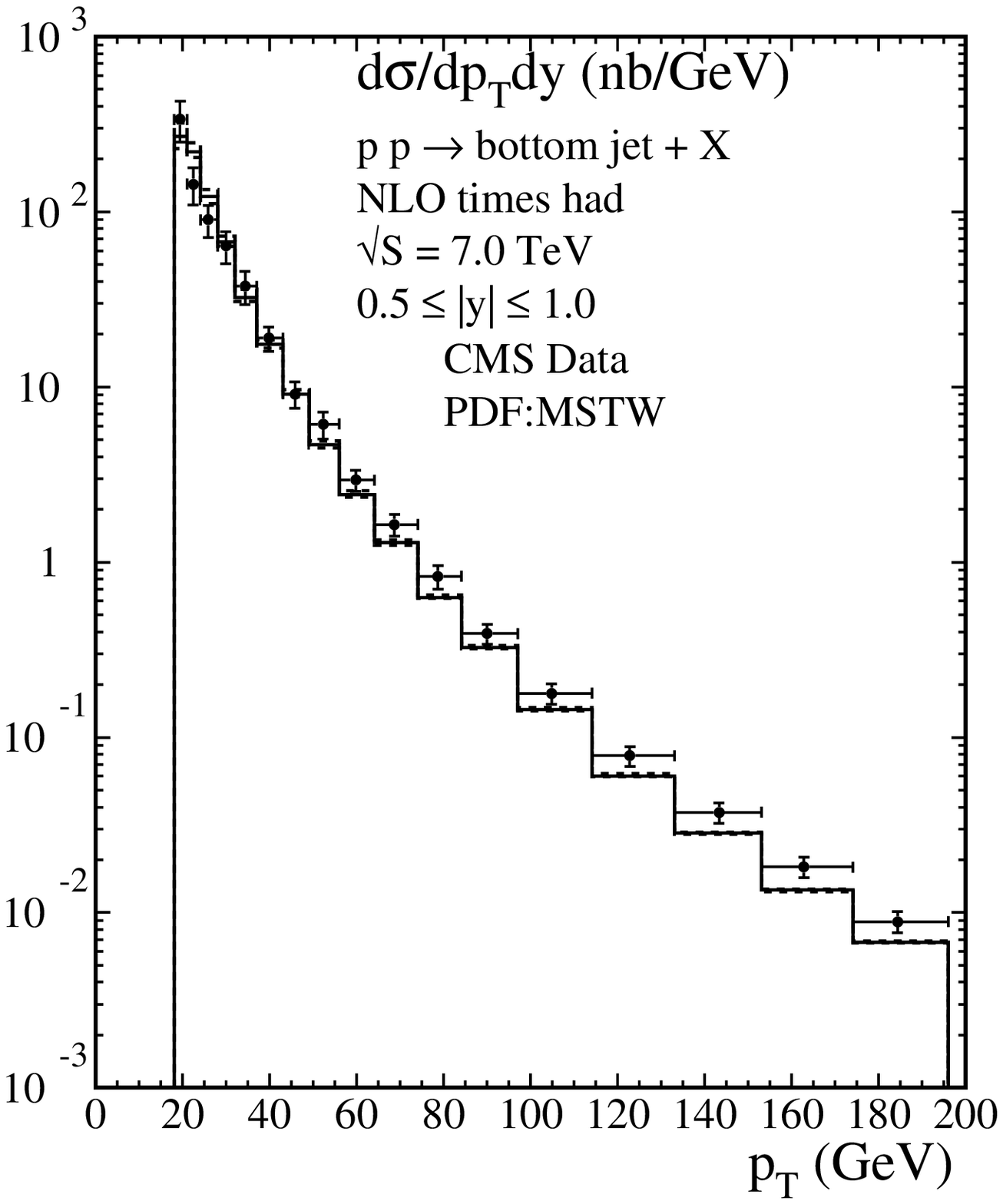}
\caption{\label{fig:5} The same caption as in Fig.~4, except for the
  rapidity region $0.5 \leq |y| \leq 1.0$.}
\end{figure*}

As already mentioned in connection with the comparison to the CDF
cross section data, the PYTHIA predictions contain not only the
hadronization corrections, but also additional parton shower
corrections, which on the other hand are also part of NLO corrections.
Of course, not all parton shower corrections would be equivalent to
our NLO corrections, since they contain additional higher order
contributions not contained in our NLO corrections. The numerical data
given to us contain both contributions, hadronization corrections and
parton shower corrections, together.

Usually the parton shower corrections contribute at small $p_T$ and
diminish for increasing $p_T$. Suppose they would be negligible at the
highest $p_T$ bin, then the hadronic corrections would still be larger
than our NLO prediction in this bin (see Fig.~4 and 5, left
frames). Since we have no information on the details of the
"non-perturbative" PYTHIA corrections, we cannot subtract the
contribution originating from the parton shower corrections to obtain
the purely hadronic corrections, etc, which are usually added to the
perturbative NLO cross sections. We expect that after subtraction of
the parton shower corrections, the agreement between our results and
the CMS data would deteriorate and we would obtain similar results as
for the MC@NLO result in Ref.~\cite{2}, although these results have an
appreciable error \cite{2, 30} due to varying the renormalization
scale, from variation of the parameters of the CTEQ PDF and from
changing the b-quark mass.

\section{Summary and Conclusions}

We have calculated the inclusive bottom-jet cross section at NLO of
QCD in the ZM-VFN scheme, i.e., with active bottom quarks in the
proton and antiproton at $\sqrt{s}=1.96$ TeV for $p\bar{p}$ and
$\sqrt{s}=7$ TeV for $pp$ collisions. Both heavy quarks are considered
massless. Our results are compared to experimental jet cross section
measurements by the CDF collaboration at the Tevatron and the CMS
collaboration at the LHC. To our surprise, the NLO cross section for
both $\sqrt{s}$ energies are much smaller than the measured cross
sections. The NLO cross sections are for most of the considered $p_T$
range even smaller than the LO cross sections, i.e. the K factors are
smaller than one. This means that the total NLO contribution, i.e.,
without the LO contribution, is negative. There exist several
possibilities to increase this contribution. Examples are:
contributions originating from intrinsic b-quark contribution to the
hadron PDF, as has been considered recently by Lyonnet et al. in
Ref.~\cite{31}, or contributions from NNLO, which might become known
in the future. Another possibility are non-perturbative hadronic
corrections which then must be much larger than have been estimated
with the PYTHIA Monte Carlo routine in Ref.~\cite{1}. This approach
has been followed in this work. It turns out that if the PYTHIA
predictions minus the LO perturbative cross section is added to our
NLO predictions, reasonable agreement with the measured b-jet cross
sections can be achieved for the CDF data as well as for the CMS
data. This approach has the problem of double counting due to the
parton shower contributions in the PYTHIA Monte Carlo approach, a
topic which we leave for investigations in the future.

\section*{Acknowledgements} 

We thank M. Butensch\"on for reading the numerical cross section values from
Ref.~\cite{1} and R Kogler and M. A. Voutilainen for communicating the PYTHIA
prediction for the CMS b-jet cross sections of Ref.~\cite{2}. IB acknowledges
support by the German Science Foundation DFG through the Collaborative Research
Centre 676 ``Particles, Strings and the Early Universe''.

\end{document}